\documentclass[english,aps,preprint,showpacs]{revtex4-1}
\usepackage[T1]{fontenc}
\usepackage[latin9]{inputenc}
\usepackage{geometry}
\usepackage{amsmath}
\usepackage{graphicx}

\makeatletter
 
 \@ifundefined{textcolor}{}
 {%
   \definecolor{BLACK}{gray}{0}
   \definecolor{WHITE}{gray}{1}
   \definecolor{RED}{rgb}{1,0,0}
   \definecolor{GREEN}{rgb}{0,1,0}
   \definecolor{BLUE}{rgb}{0,0,1}
   \definecolor{CYAN}{cmyk}{1,0,0,0}
   \definecolor{MAGENTA}{cmyk}{0,1,0,0}
   \definecolor{YELLOW}{cmyk}{0,0,1,0}
 }

\makeatother

\usepackage{babel}
\begin{document}

\title{Stress-induced patterns in ion-irradiated Silicon: a model based
on anisotropic plastic flow}

\author{Scott A. Norris}

\affiliation{Department of Mathematics\\
Southern Methodist University, Dallas, TX 75275}

\email{snorris@smu.edu}

\date{July 23, 2012}

\begin{abstract}
We present a model for the effect of stress on thin amorphous films
that develop atop ion-irradiated silicon, based on the mechanism of
ion-induced anisotropic plastic flow. Using only parameters directly
measured or known to high accuracy, the model exhibits remarkably
good agreement with the wavelengths of experimentally-observed patterns,
and agrees qualitatively with limited data on ripple propagation speed.
The predictions of the model are discussed in the context of other
mechanisms recently theorized to explain the wavelengths, including
extensive comparison with an alternate model of stress.
\end{abstract}
\maketitle

\section{Introduction}

Among the many nanoscale patterns that have been observed on ion-irradiated
surfaces \cite{chan-chason-JAP-2007,frost-etal-APA-2008}, the discovery
of hexagonal arrays of high-aspect ratio dots on Ar\textsuperscript{+}-irradiated
GaSb \cite{facsko-etal-SCIENCE-1999} has sparked a flurry of experimental
and theoretical study into spontaneous pattern formation as a potential
route to ``bottom-up'' fabrication of nanoscale devices. A growing
body of evidence increasingly suggests that highly-ordered structures
occur only for targets containing more than one material \cite{ozaydin-etal-APL-2005,macko-etal-NanoTech-2010,bradley-shipman-PRL-2010}.
Although early observations of dot formation on pure Silicon \cite{gago-etal-APL-2006,ziberi-etal-APL-2008,madi-etal-2008-PRL}
sparked a series of proposed explanations for the ordered structures
\cite{facsko-etal-PRB-2004,davidovitch-etal-PRB-2007}, none turned
out to be viable \cite{bradley-PRB-2011a,norris-PRB-2012-viscoelastic-normal}:
ultimately, dots disappeared when impurities and geometric artifacts
were carefully removed \cite{madi-aziz-ASS-2012}, and reappeared
upon their systematic re-introduction \cite{ozaydin-etal-APL-2005,ozaydin-etal-JVSTB-2008,ozaydin-ludwig-JPCM-2009,zhang-etal-NJoP-2011}.
As a result, much recent attention has focused on theories of irradiated
binary materials \cite{shenoy-chan-chason-PRL-2007,bradley-shipman-PRL-2010,shipman-bradley-PRB-2011,bradley-PRB-2011c,bradley-JAP-2012,abrasonis-morawetz-PRB-2012,norris-PRB-2012-chemical-instability}. 

Despite these results on the formation of ordered structures, the
study of monatomic targets remains important because of commonalities
between binary and monatomic systems, and the relative simplicity
of the latter. In particular, because it is readily amenable to molecular
dynamics simulation, and its near-surface region is amorphous under
ion bombardment, noble-gas ion irradiation of silicon has been an
important system for comparison between experiment, theory, and simulation
of pure materials. This allows rapid development and testing of theories
on the basic physical processes of ion irradiation, which occur on
two time scales. On timescales of individual ion impacts, spanning
$\sim10^{-9}$ sec and described as the \emph{prompt regime}, there
is erosion of some target atoms away from the target \cite{sigmund-PR-1969,sigmund-JMS-1973,bradley-harper-JVST-1988},
and redistribution of others to new locations \cite{carter-vishnyakov-PRB-1996,davidovitch-etal-PRB-2007}.
On much longer timescales associated with kinetic relaxation, spanning
around $\sim10^{2}$ sec and denoted the \emph{gradual regime}, there
is the possibility of surface diffusion \cite{bradley-harper-JVST-1988},
stress buildup \cite{volkert-JAP-1991}, and viscous flow \cite{umbach-etal-PRL-2001}.
The understanding of these basic processes, which remain important
in binary materials, continues to evolve in important ways, bringing
closer the goal of a theory developed enough to make accurate predictions
on irradiation-induced structure formation.

In this paper, we investigate the effect on morphology evolution
of stress buildup and relaxation in the thin film of material affected
by ion irradiation. We choose a different approach than a recent series
of papers on this topic \cite{cuerno-etal-NIMB-2011,castro-cuerno-ASS-2012,castro-etal-arXiv-2012-solid-flow-drives},
which cast the role of stress in terms of classical fluid dynamics
results on gravity-driven flows \cite{oron-davis-bankoff-RMP-1997};
the relationship between that model and our own will be discussed
in detail in Section \ref{sec: castro-comparison}. Here, instead,
we draw analogy with the large literature on ion-induced stress for
irradiation at energies in the MeV range, where incoming ions are
primarily slowed by electronic stopping \cite{trinkaus-ryazanov-PRL-1995-viscoelastic,trinkaus-NIMB-1998-viscoelastic,snoeks-etal-JAP-1995,van-dillen-etal-APL-2001-colloidal-ellipsoids,van-dillen-etal-PRB-2005-viscoelastic-model,otani-etal-JAP-2006}.
 There, incoming ions induce an anisotropic deformation in the material,
leading to stress buildup. More recently, this effect has also been
observed during irradiation in the keV range \cite{van-dillen-etal-APL-2003-colloidal-ellipsoids},
and the mathematical form of the model has been successfully used
to describe a number of phenomena at even lower energies where nuclear
stopping is dominant \cite{kim-etal-JAP-2006-stressed-keV-films,george-etal-JAP-2010}.
Based on these results, we here apply the model formally to pattern
formation at low energies. We find that our model is able to predict
the observed wavelength to remarkable accuracy without the need to
estimate any unknown parameter values. We also find it makes predictions
on the ripple propagation speed that are consistent with experimental
observations.

\section{Model}

We consider the morphological evolution of the top layer of a monatomic
target, irradiated by noble gas ions at an incidence angle $\theta$.
The ions penetrate a certain average distance into the solid before
initiating a collision cascade of atomic displacements. Over many
impacts, the accumulated damage induced by these cascades amorphizes
a thin film of material atop the target, with a crystalline/amorphous
boundary at the bottom the film, and a free boundary at the top of
the film. We construct a co-ordinate system in which the (planar)
target is perpendicular to the $z$-axis, with $z=h\left(x,y\right)$
tracking the top free boundary, and $z=g\left(x,y\right)$ tracking
the bottom boundary. We will then consider the effect of stress on
the stability of this configuration. In this paper, we will focus
exclusively on the effect of stress, and therefore neglect the prompt
regime of sputtered atoms \cite{sigmund-PR-1969,bradley-harper-JVST-1988},
or those redistributed to new locations \cite{carter-vishnyakov-PRB-1996,davidovitch-etal-PRB-2007,norris-etal-NCOMM-2011}.

\subsection{Constitutive Law}

\paragraph*{Basic Material Properties.}

It has been established for some time that the incoming ions impart
an effective fluidity to the amorphous film that is many orders of
magnitude larger than that of bulk amorphous silicon \cite{umbach-etal-PRL-2001}.
So any treatment of the film must include viscous effects. Proposals
in past years have suggested the additional consideration of elastic
effects \cite{davidovitch-etal-PRB-2007,madi-etal-2008-PRL}, and
a constitutive law including all such effects is \cite{otani-etal-JAP-2006}
\begin{equation}
\dot{\mathbf{E}}=\frac{1}{2\eta}\mathbf{T}+\frac{1}{2G}\dot{\mathbf{T}}+\frac{1}{9B}\frac{D}{Dt}\left(\text{tr}\mathbf{T}\right)\mathbf{I}+\dot{\mathbf{E}}_{B}\label{eqn: otani-cr}
\end{equation}
Here the first three terms contributing to the linear rate-of-strain
tensor 
\begin{equation}
\dot{\mathbf{E}}=\frac{1}{2}\left(\nabla\mathbf{v}+\nabla\mathbf{v}^{T}\right)\label{eqn: rate-of-strain}
\end{equation}
 constitute a standard linear model for a Maxwell fluid \cite{malvern-1977},
where $\mathbf{T}$ is the stress tensor, $\eta$ is the viscosity,
$G$ is the shear modulus, and $B$ is the bulk modulus. The final
term on the right, $\dot{\mathbf{E}}_{B}$, is a contribution to this
rate of strain due the ion beam. However, it has been argued elsewhere
that, in a variety of relevant regimes, the contribution of elasticity
should be negligible \cite{george-etal-JAP-2010,norris-PRB-2012-viscoelastic-normal}.
Hence, we take the limit of a purely viscous ($G\to\infty$) and incompressible
($B\to\infty$) fluid. In the incompressible limit, however, a hydrostatic
pressure term must also be included, leading to the simpler constitutive
law we will consider for the remainder of this work,
\begin{equation}
\mathbf{T}=-p\mathbf{I}+2\eta\left(\dot{\mathbf{E}}-\dot{\mathbf{E}}_{B}\right),\label{eqn: newton-plus-stress-CR}
\end{equation}
which is nearly equivalent to a Newtonian fluid, except for the addition
of the constant contribution $-2\eta\dot{\mathbf{E}}_{B}$ to the
stress.

\paragraph*{Effect of Stress.}

We now turn to the modeling of stress. Recent work by Castro, Cuerno,
and co-workers has appealed to analogy with familiar concepts in fluid
dynamics by assuming that the microscopic mechanism stress generation
can be coarse-grained into an effective body force \cite{cuerno-etal-NIMB-2011,castro-cuerno-ASS-2012}.
We will discuss that approach in Section \ref{sec: castro-comparison},
but will here take a different approach, attempting to connect more
directly with the microscopic process. At that scale, MD simulation
has shown that each impact significantly redistributes the target
silicon atoms to new locations \cite{moseler-etal-SCIENCE-2005,kalyanasundaram-etal-APL-2008,norris-etal-NCOMM-2011},
gradually increasing the magnitude of a compressive stress to a saturated
state \cite{kalyanasundaram-etal-AM-2006}, which is also observed
experimentally \cite{madi-thesis-2011}. Each impact thus induces
a direct deformation of the material, suggesting that the effect of
the beam be incorporated directly into the constitutive relationship
between stress and strain, in a way that depends linearly on the total
fluence.

A model with exactly these properties has already been developed to
describe anisotropic plastic flow during high-energy ion irradiation
in the regime of electronic stopping, where rapid thermal cycling
due to ion impacts leads to a stress-free strain rate of the form
\cite{trinkaus-ryazanov-PRL-1995-viscoelastic,trinkaus-NIMB-1998-viscoelastic,van-dillen-etal-PRB-2005-viscoelastic-model,otani-etal-JAP-2006}
\begin{equation}
\dot{\mathbf{E}}_{B}=fA\mathbf{D}\left(\theta\right).\label{eqn: strain-rate-beam}
\end{equation}
Here, $f$ is the ion flux, $A$ is a measure of the magnitude of
strain induced per ion, and $\mathbf{D}$ describes the (anisotropic
and angle-dependent) shape of that strain. The \emph{mechanism} of
behind Eqn.(\ref{eqn: strain-rate-beam}) is of course not directly
applicable in the nuclear stopping regime. However, the \emph{phenomenon}
anisotropic plastic flow has been observed even in the nuclear stopping
regime \cite{van-dillen-etal-APL-2003-colloidal-ellipsoids}, and
the \emph{mathematical form} of the governing equations has been applied
successfully to describe various phenomena at low energies \cite{kim-etal-JAP-2006-stressed-keV-films,george-etal-JAP-2010}.
In this spirit, we have argued elsewhere \cite{norris-PRB-2012-viscoelastic-normal}
that, for $\theta=0$, simple symmetry arguments lead almost immediately
to the form
\begin{equation}
\mathbf{D}\left(0\right)=\left[\begin{array}{ccc}
1 & 0 & 0\\
0 & 1 & 0\\
0 & 0 & -2
\end{array}\right]\label{eqn: normal-D}
\end{equation}
regardless of the underlying mechanism. From here, it is natural to
add angle-dependence via the matrix
\begin{equation}
\mathbf{R}\left(\theta\right)=\left[\begin{array}{ccc}
\cos\theta & 0 & -\sin\left(\theta\right)\\
0 & 1 & 0\\
\sin\left(\theta\right) & 0 & \cos\left(\theta\right)
\end{array}\right],
\end{equation}
describing rotation about the $y$-axis, which leads to
\begin{equation}
\mathbf{D}\left(\theta\right)=\mathbf{R}\left(-\theta\right)\mathbf{D}\left(0\right)\mathbf{R}\left(\theta\right)=\left[\begin{array}{ccc}
\frac{3}{2}\cos\left(2\theta\right)-\frac{1}{2} & 0 & \frac{3}{2}\sin\left(2\theta\right)\\
0 & 1 & 0\\
\frac{3}{2}\sin\left(2\theta\right) & 0 & -\frac{3}{2}\cos\left(2\theta\right)-\frac{1}{2}
\end{array}\right].\label{eqn: APF-tensor}
\end{equation}
Combining (\ref{eqn: newton-plus-stress-CR}), (\ref{eqn: strain-rate-beam}),
we obtain the final form of the constitutive relation, 
\[
\mathbf{T}=-p\mathbf{I}+2\eta\left(\dot{\mathbf{E}}-fA\mathbf{D}\right),
\]
with $\mathbf{D}\left(\theta\right)$ given by Eqn.(\ref{eqn: APF-tensor}).

\subsection{Governing Equations}

The governing equations are obtained following standard continuum
analysis in the limit of a highly viscous film. In the bulk, conservation
of mass in the incompressible limit gives
\begin{equation}
\nabla\cdot\mathbf{v}=0,\label{eqn: incompressibility}
\end{equation}
the incompressibility condition on the velocity field $\mathbf{v}$.
In addition, conservation of momentum takes the form
\begin{equation}
\rho\left(\frac{\partial\mathbf{v}}{\partial t}+\mathbf{v}\cdot\nabla\mathbf{v}\right)=\nabla\cdot\mathbf{T}+\mathbf{f},\label{eqn: conservation-of-momentum}
\end{equation}
where the left hand side contains the density $\rho$ and the material
acceleration, and the right hand side lists forces $\nabla\cdot\mathbf{T}$
due to internal stresses, and $\mathbf{f}$ due to long-range body
forces. From here, because of the high viscosity of the ion-irradiated
film, accelerations in the left-hand side are small and can be neglected,
and we do not consider the effect of any body forces. Hence, we arrive
at Stokes equations: 
\begin{equation}
\nabla\cdot\mathbf{T}=-\nabla p+\eta\nabla^{2}\mathbf{v}=0.\label{eqn: stokes-equations}
\end{equation}

It remains to apply boundary conditions. At the amorphous/crystalline
boundary $z=g\left(x,y\right)$ we have the combined ``no-penetration''
and ``no-slip'' conditions for a viscous fluid;
\begin{equation}
\mathbf{v}=\mathbf{0}.\label{eqn: no-slip-or-penetration-BC}
\end{equation}
Meanwhile, allowing for the effect of surface tension on a viscous
fluid, the stress balance at the free boundary $z=h\left(x,y\right)$
can be written
\begin{equation}
\mathbf{T}\cdot\mathbf{n}=-\gamma\kappa\mathbf{n}\label{eqn: surface-stress-BC}
\end{equation}
while the kinematic condition relating bulk velocity \textbf{$\mathbf{v}$}
to the normal surface velocity at the interface, $v_{I}$, is 
\begin{equation}
v_{I}=\mathbf{v}\cdot\mathbf{n}\label{eqn: kinematic-condition}
\end{equation}
We see that the effect of the beam, through the angle-dependent tensor
$\mathbf{D}\left(\theta\right)$, appears mathematically only at the
free boundary in Equation (\ref{eqn: surface-stress-BC}), where it
alters the conditions necessary for stress balance.

\section{Analysis}

\subsection{Steady Solution}

We first look for a steady state ($\partial/\partial t\to0$) consisting
of a flat, uniform film occupying the space between $z=g_{0}$ and
$z=h_{0}$, with pressure $p_{0}$ and velocity $\mathbf{v}_{0}$.
For convenience we choose $g_{0}=0$. Symmetry considerations greatly
simplify the calculations -- the assumption of uniformity implies
translational symmetry in $x$ and $y$, which limits the steady pressure
and velocity field to be functions of $z$ at most: $p_{0}\left(z\right)$
and $\mathbf{v}_{0}\left(z\right)$. Furthermore, although off-normal
ion incidence from the negative $x$- direction may break reflection
symmetry in $x$, we enforce reflection symmetry in $y$, which prohibits
a velocity component in the $y$-direction. These considerations immediately
limit the form of the steady velocity field $\mathbf{v}_{0}$ to 
\begin{equation}
\mathbf{v}_{0}\left(z\right)=\left(u_{0}\left(z\right),\,0,\, w_{0}\left(z\right)\right)^{T}.\label{eqn: steady-velocity-form}
\end{equation}
Comparing Eqn. (\ref{eqn: steady-velocity-form}) to the incompressibility
condition (\ref{eqn: incompressibility}) and the no-penetration condition
(\ref{eqn: no-slip-or-penetration-BC}), we first see that
\begin{equation}
w_{0}\left(z\right)=0.
\end{equation}
Next, the $z$-component of the Stokes equations (\ref{eqn: stokes-equations})
show the steady pressure to be a constant
\begin{equation}
p_{0}\left(z\right)=a,
\end{equation}
while the $x$-component of (\ref{eqn: stokes-equations}) restricts
$u_{0}\left(z\right)$ to at most the linear form 
\begin{equation}
u_{0}\left(z\right)=b+cz.
\end{equation}
Finally, application of the remaining boundary conditions gives us
$a$, $b$, and $c$. The no-slip condition (\ref{eqn: no-slip-or-penetration-BC}),
applied at $z=g_{0}=0$, requires $b=0$, while the stress balance
condition (\ref{eqn: surface-stress-BC}), applied at $z=h_{0}$,
yields $a=\eta fA\left[3\cos\left(2\theta\right)+1\right]$ and $c=3fA\sin\left(2\theta\right)$.
In summary, the steady solution is thus
\begin{equation}
\begin{aligned}p_{0} & =\eta fA\left[3\cos\left(2\theta\right)+1\right]\\
\mathbf{v}_{0} & =\left(3fA\sin\left(2\theta\right)z,\,0,\,0\right)^{T}\\
\mathbf{T}_{0} & =-6\eta fA\left[\begin{array}{lcr}
\cos\left(2\theta\right) & 0 & \qquad0\\
0 & \cos^{2}\left(\theta\right) & \qquad0\\
0 & 0 & \qquad0
\end{array}\right]
\end{aligned}
,\label{eqn: steady-state}
\end{equation}
consisting of a uniform pressure and downbeam shear that both depend
on the irradiation angle-of-incidence $\theta$. The pressure is maximum
at $\theta=0^{\circ}$ and monotonically decreases at higher angles,
whereas the shear flow is maximal at $\theta=45^{\circ}$ and decays
to zero for $\theta=\left\{ 0^{\circ},\,90^{\circ}\right\} $. We
note in particular the form of the steady stress when $\theta=0$;
a symmetric, biaxial compressive stress with no vertical component.

\subsection{Linear Stability}

We now perform a linear stability analysis -- e.g., we consider the
evolution of the system when the top boundaries is subjected to an
infinitesimal, sinusoidal perturbation in the $x$- and $y$- directions.
This leads to solutions that can be expressed in terms of \emph{normal
mode}s having the form
\begin{equation}
\left[\begin{array}{c}
g\\
h\\
p_{\phantom{0}}\\
\mathbf{v}_{\phantom{0}}
\end{array}\right]=\left[\begin{array}{c}
0\\
h_{0}\\
p_{0}\\
\mathbf{v}_{0}\left(z\right)
\end{array}\right]+\varepsilon\left[\begin{array}{c}
g_{1}\\
h_{1}\\
p_{1}\left(z\right)\\
\mathbf{v}_{1}\left(z\right)
\end{array}\right]e^{i\left(k_{1}x+k_{2}y\right)+\sigma t}.\label{eqn: perturbations}
\end{equation}
Here we do not assume that the bottom boundary remains flat. Instead,
as the top boundary changes shape, the zone of material affected by
collision cascades changes with it. As a simplest approximation to
this behavior, based on the observed weak dependency of film thickness
on angle \cite{madi-2009-MRS-castro-response}, we choose 
\[
g_{1}=h_{1}
\]
 indicating that the bottom boundary is simply a vertical translation
of the top boundary, as depicted in Figure~\ref{fig: ripple-flow-mechanism}.
With this definition, the forms (\ref{eqn: perturbations}) are then
inserted inserted into the governing equations (\ref{eqn: stokes-equations})-(\ref{eqn: kinematic-condition}),
and terms are kept only to leading order in the vanishingly-small
parameter $\varepsilon$. Solution proceeds in three steps, as shown
in the Appendix. First, the linearized version of the incompressibility
condition (\ref{eqn: incompressibility}) and Stokes equations (\ref{eqn: stokes-equations})
are solved to give a general solution for the pressure and velocity
fields in the bulk. Second, the linearized boundary conditions (\ref{eqn: no-slip-or-penetration-BC})
and (\ref{eqn: surface-stress-BC}) are applied to determine the resulting
integration constants. Last, having uniquely determined the pressure
and velocity fields, the application of the kinematic condition (\ref{eqn: kinematic-condition})
leads directly to the \emph{dispersion relation} $\sigma\left(k_{1},\, k_{2}\right)$
governing the evolution of each mode. We find

\begin{equation}
\begin{aligned}\sigma\left(k_{1},k_{2}\right) & =-6fA\frac{\cos\left(2\theta\right)\left(h_{0}k_{1}\right)^{2}+\cos^{2}\left(\theta\right)\left(h_{0}k_{2}\right)^{2}}{1+2Q^{2}+\cosh\left(2Q\right)}-3fA\sin\left(2\theta\right)i\left(h_{0}k_{1}\right)\\
 & -3fA\sin\left(2\theta\right)i\left(h_{0}k_{1}\right)\left\{ \frac{2\cosh\left(Q\right)\left[Q^{2}+\sinh^{2}\left(Q\right)\right]}{1+2Q^{2}+\cosh\left(2Q\right)}-\cosh\left(Q\right)\right\} \\
 & -\frac{\gamma}{2\eta h_{0}}\frac{Q\left(\sinh\left(2Q\right)-2Q\right)}{1+2Q^{2}+\cosh\left(2Q\right)}
\end{aligned}
.\label{eqn: full-dispersion-relation}
\end{equation}
where the dimensionless wavevector $Q=h_{0}\sqrt{k_{1}^{2}+k_{2}^{2}}$.
Here the first line corresponds to the effect of anisotropic plastic
flow operating on the bulk film, the second line corresponds to the
effect (under said flow) of the non-planar bottom boundary $g_{1}\left(x,y\right)=h_{1}\left(x,y\right)$,
and the third line corresponds to the well-studied (isotropic) effect
of surface leveling under surface tension (see Ref. \cite{orchard-ASR-1962}).

A commonly-employed simplification of Eqn.(\ref{eqn: full-dispersion-relation})
occurs in the long-wavelength limit $Q\ll1$ of Eqn.(\ref{eqn: full-dispersion-relation}),
in which wavelength is much longer than the film depth. In this limit
the leading-order contributions of each line in Eqn.(\ref{eqn: full-dispersion-relation})
reduce to 
\begin{equation}
\begin{aligned}\sigma & \approx-3fA\left[\cos\left(2\theta\right)\left(k_{1}h_{0}\right)^{2}+\cos^{2}\left(\theta\right)\left(k_{2}h_{0}\right)^{2}\right]\\
 & -\frac{9}{2}fA\sin\left(2\theta\right)i\left(k_{1}h_{0}\right)Q^{2}\\
 & -\frac{\gamma}{3\eta h_{0}}Q^{4}
\end{aligned}
.\label{eqn: longwave-dispersion-relation}
\end{equation}
This approximation is useful by nature of being much simpler than
Eqn.(\ref{eqn: full-dispersion-relation}). However, despite being
commonly employed, this assumption is less commonly verified against
experimental data. For instance, in the experiments of Madi \cite{madi-etal-2008-PRL,madi-etal-JPCM-2009},
widely used as a point of comparison with theory, the film thickness
$h_{0}\approx3\,\mathrm{nm}$, while wavenumbers are as large as $k\approx2\pi/\left(20\,\mathrm{nm}\right)$,
giving $Q\approx1$. Hence, while we include Eqn.(\ref{eqn: longwave-dispersion-relation})
for reference and use in qualitative discussion, numerical predictions
will be based on the full Eqn.(\ref{eqn: full-dispersion-relation}).

\subsection{Discussion of Results}

\paragraph*{Stability.}

\begin{figure}

\begin{centering}
\includegraphics[width=5in]{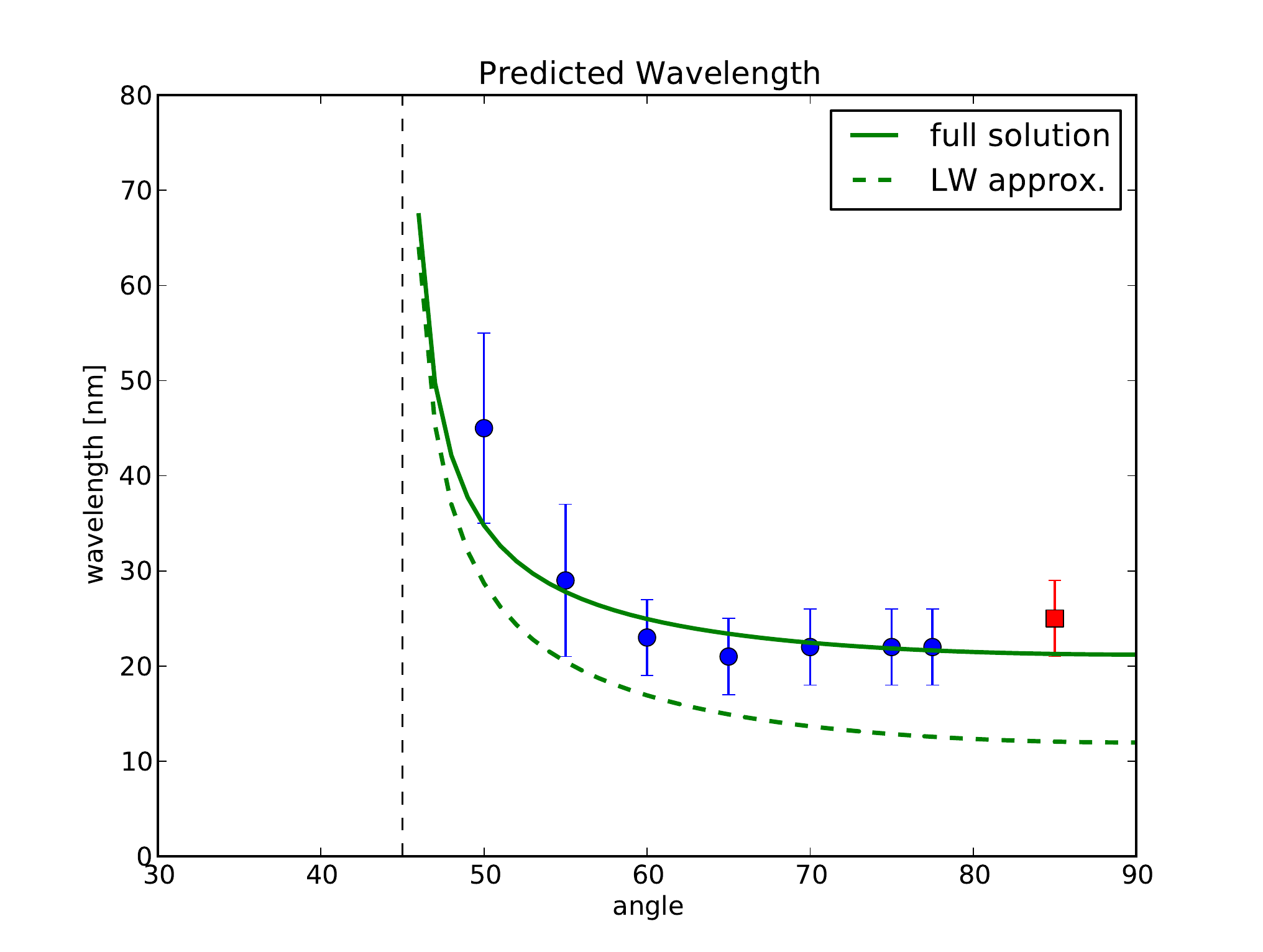}\caption{(color online) Wavelength as predicted by numerically maximizing the
full dispersion relation (\ref{eqn: full-dispersion-relation}) over
$Q$ (solid line), and under the longwave approximation (\ref{eqn: longwave-dispersion-relation})
(dashed line). Given the fixed parameter values indicated in the text,
the full dispersion relation predicts the observed wavelengths of
parallel-mode ripples (circles) to surprising accuracy, although it
does not predict the rotation of the ripples to perpendicular-mode
(square) at high angles. Following the discussion the text, we observe
that the wavelengths predicted by the longwave approximation differ
significantly from those predicted by the full dispersion relation.}

\par\end{centering}

\label{fig: predicted-wavelength}

\end{figure}

The real terms 
\[
r\left(k\right)=\mathrm{Re}\left[\sigma\left(k\right)\right]
\]
 from Eqn.(\ref{eqn: longwave-dispersion-relation}) describe the
\emph{growth rate} of perturbations to the flat steady state in terms
of their wavenumber $k$. The value of $k$ that maximizes $r\left(k\right)$
- denoted $k^{*}$ - is called the \emph{most unstable mode}; if $\sigma\left(k^{*}\right)<0$,
then all modes decay and the steady solution is \emph{stable} , whereas
if $\sigma\left(k^{*}\right)>0$, then at least some modes near $k^{*}$
grow, and the steady solution is unstable. From the full dispersion
relation (\ref{eqn: full-dispersion-relation}), we see that a positive
band of unstable wavenumbers exists if $\theta>45^{\circ}$. In the
longwave approximation (\ref{eqn: longwave-dispersion-relation}),
the most unstable wavelength is given by 
\begin{equation}
\lambda^{*}\approx2\pi\sqrt{\frac{2\gamma h}{9fA\eta\cos\left(2\theta\right)}}\approx2\pi\sqrt{\frac{4\gamma h}{3\left|\mathbf{T}_{0}\left(0\right)\right|\cos\left(2\theta\right)}}\label{eqn: wavelength-prediction}
\end{equation}
where the denominator inside the radical is identified as a multiple
of the steady in-plane stress at normal incidence, as given by Eqn.(\ref{eqn: steady-state}).
This is convenient, because it allows replacing the estimation of
the infrequently-measured ion-enhanced viscosity $\eta$, and the
stress-per-ion parameter $A$, with an experimental observation of
the steady stress at normal incidence. Hence, we can compare existing
measurements of the steady stress and the wavelength for the same
material under identical irradiation conditions. It has already been
shown by Madi that at $250$ eV, $h\approx3\,\mathrm{nm}$ and $\left|\mathbf{T}_{0}\left(0\right)\right|\approx1.5\,\mathrm{GPa}$
\cite{madi-thesis-2011}, while in the same chamber at the same energy,
wavelengths given by Figure \ref{fig: predicted-wavelength} were
observed \cite{madi-etal-2008-PRL,madi-etal-JPCM-2009,madi-aziz-ASS-2012}.
With the additional value $\gamma=1.36\,\text{J/m}$ \cite{vauth-mayr-PRB-2007,eaglesham-etal-PRL-1993},
we can compare the stress measurements to the observed wavelengths.
We find that the full dispersion relation predicts a remarkably good
fit to the data. Unsurprisingly given the size of $Q$, the longwave
approximation does not do nearly as well. 

\paragraph*{Ripple Velocity.}

The imaginary terms 
\[
\omega\left(k\right)=-\mathrm{Im}\left[\sigma\left(k\right)\right]
\]
 from Eqn.(\ref{eqn: longwave-dispersion-relation}) (note the conventional
minus sign) describe the \emph{propagation rate} of perturbations
to the flat steady state because, from the original linear ansatz,
\begin{equation}
\exp\left(\sigma t+ikx\right)=\exp\left(rt+ik(x-\frac{\omega}{k}t)\right).\label{eqn: phase-velocity}
\end{equation}
Equation (\ref{eqn: phase-velocity}) reveals that, in a stationary
``lab'' frame of reference, individual ripples of wavelength $k$
propagate with a speed 

\[
V_{\text{phase}}=\frac{\omega\left(k\right)}{k}
\]
usually called the \emph{phase velocity}. Comparison with (\ref{eqn: longwave-dispersion-relation})
reveals that, for our problem, 
\begin{equation}
V_{\text{phase}}=3fAh_{0}\sin\left(2\theta\right)\left\{ \frac{2\cosh\left(Q\right)\left[Q^{2}+\sinh^{2}\left(Q\right)\right]}{1+2Q^{2}+\cosh\left(2Q\right)}-\cosh\left(Q\right)\right\} \approx\frac{9}{2}fAh_{0}\sin\left(2\theta\right)Q^{2}.\label{eqn: ripple-velocity}
\end{equation}
Although limited experimental data on ripple velocities is available,
what reports exist indicate that - contrary to early seminal theory
concerned only with erosion \cite{bradley-harper-JVST-1988} - ripples
always propagate in the direction of the ion beam \cite{alkemade-PRL-2006,gnaser-etal-NIMB-2012-ripple-speed}.
Both our longwave result (\ref{eqn: ripple-velocity}), and also the
velocity resulting from the full dispersion relation (\ref{eqn: full-dispersion-relation}),
are uniformly positive for all values of $\theta$ or $Q$, consistent
with these observations.

\begin{figure}
\begin{centering}
\includegraphics[width=5in]{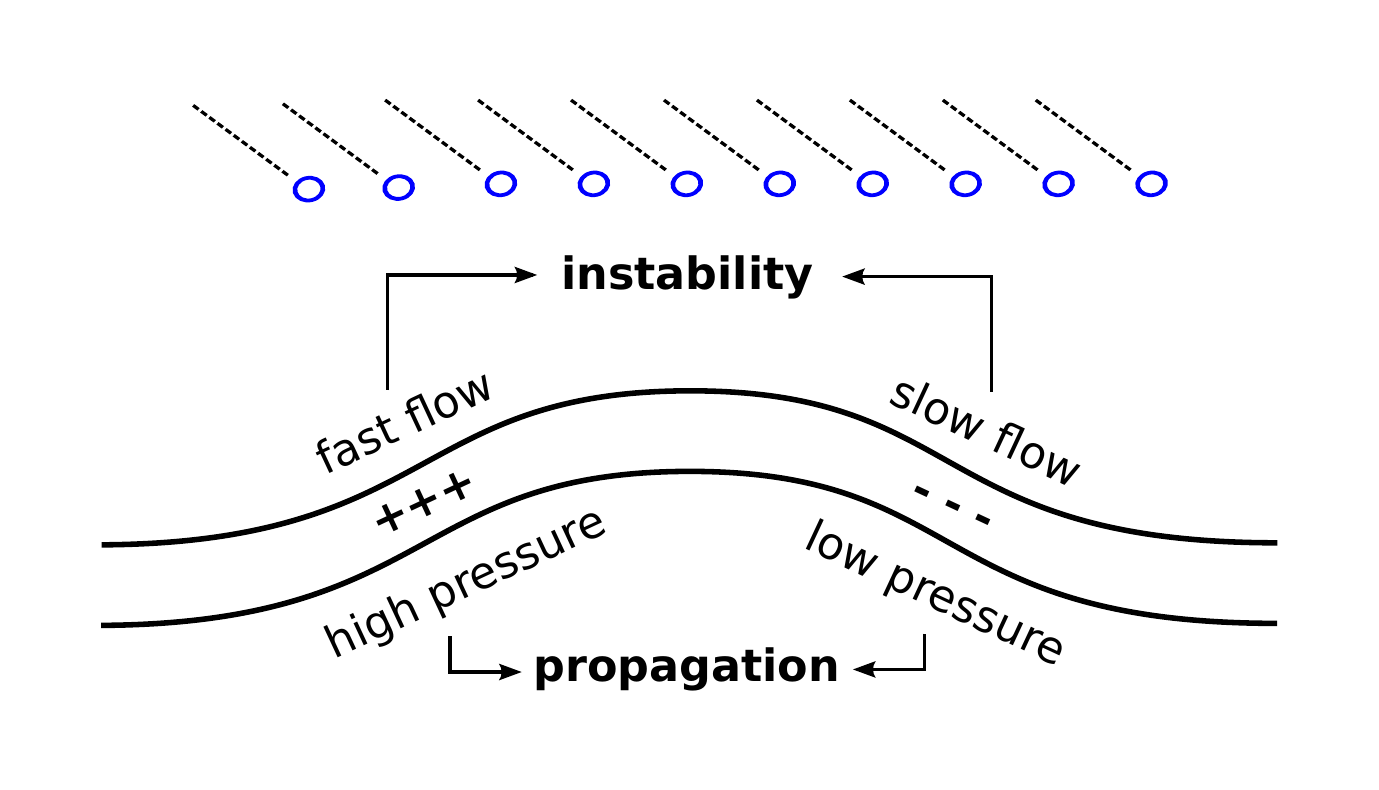}\caption{Schematic of the system, illustrating the intuitive understanding
of ripple growth and propagation.}

\par\end{centering}

\label{fig: ripple-flow-mechanism}
\end{figure}

\paragraph*{Intuitive Explanation of Ripple Growth and Propagation.}

Both the instability and translation mechanisms can be understood
intuitively in terms of simple differences in the steady shear flow
and pressure across a ripple. The instability occurs when the net
shear flow on the uphill side exceeds the net flow on the downhill
side. We recall that the flow rate has a maximum when the local angle
of incidence is $45^{\circ}$, and decreases monotonically for as
the local angle continues to increase. So, when $\theta>45$, then
uphill slopes (with angle closer to $45^{\circ}$) have greater flow
than downhill slopes (with higher angle), leading to mass accumulation
at the top of the ripple. Similarly, ripple translation occurs because
the steady pressure is a monotonically decreasing function of incidence
angle. Hence, for any angle, the pressure in the uphill side exceeds
that in the downhill side, leading to flow from uphill to downhill
slopes, as indicated in Figure \ref{fig: ripple-flow-mechanism}.
We note that this ripple propagation mechanism is almost entirely
driven by the nonplanar nature of the bottom boundary; if the bottom
boundary were instead flat, the only contribution to the phase velocity
would come from the last term of the first line of (\ref{eqn: full-dispersion-relation}),
which indicates simple translation along with the shearing film. However,
when the translated bottom boundary is included, it exactly cancels
this simple shear in the long-wavelength limit, and the resulting
ripple propagation is entirely due to differences in pressure across
a ripple.

\paragraph*{Relationship to thermal stress models.}

The contribution (\ref{eqn: ripple-velocity}) to the ripple velocity,
and indeed several aspects of the analysis used to attain it, bears
a notable resemblance to the work of Alkemade on ripple propulsion
\cite{alkemade-PRL-2006}. There, the author proposes a mechanism
of ripple propagation that also is caused by variations in the stress,
but in which the stress itself is caused by thermal expansion. Briefly,
larger relative depth of the uphill slopes causes higher sustained
temperatures there, leading to greater thermal expansion and thus
greater thermal stress. The problem is that at the scale of the ripple
size ($\sim20-50\text{ nm}$), thermal diffusion is so rapid that
no meaningful temperature differential can be sustained. Our model
can be thought of as an alternative to this mechanism, where the source
of stress is purely mechanical, requiring no thermal gradients, and
variation appears naturally as a result of the different slopes inherent
to a ripple structure.

\section{Relationship to Effective Body Force Models\label{sec: castro-comparison}}

We will now attempt to give a comprehensive comparison of our theory
with another recent model of stress that appeals to analogies with
fluid dynamics \cite{cuerno-etal-NIMB-2011,castro-cuerno-ASS-2012,castro-etal-arXiv-2012-solid-flow-drives}
by rendering the effect of stress as an ``effective body force''
(EBF) which acts throughout the amorphous film. The bulk of the the
EBF model is developed in Ref.\cite{castro-cuerno-ASS-2012}. There,
the effect of the beam is posited to lie within the constitutive relation
via
\begin{equation}
T_{ij}=-p\delta_{ij}+\mu\left(\partial_{i}v_{j}+\partial_{j}v_{i}\right)+T_{ij}^{s},\label{eqn: castro-constitutive}
\end{equation}
which is effectively identical to our Eqn.(\ref{eqn: otani-cr}),
and similar to work dating at least to Ref.\cite{otani-etal-JAP-2006}.
However, the authors never explicitly define the tensor $T_{ij}^{s}$
describing the effect of the beam, instead proposing only that its
divergence takes the form 
\begin{equation}
\nabla\cdot\mathbf{T}^{s}=\mathbf{b}=\mathbf{f}_{E}\Psi\left(\theta-\gamma\right)\label{eqn: effective-body-force}
\end{equation}
where the authors ``define $\mathbf{b}=\nabla\cdot\mathbf{T}^{s}$
as a body force acting in the bulk of the fluid layer.'' Although
the incoming ions undoubtedly exert a force on the film as they are
slowed down by the film, that force can be shown to be vanishingly
small \cite{madi-2009-MRS-castro-response}, and Eqn.(\ref{eqn: effective-body-force})
is not suggested to represent any actual physical force. Instead,
$\mathbf{f}_{E}$ is proposed to contain ``the coarse-grained information
about the effect of the residual stress created in the target ...
{[}with{]} dimensions of a gradient of stress,'' and $\Psi\left(\theta-\gamma\right)$
to encode dependence upon the local angle of incidence $\theta-\gamma$.
This is an interesting effort to encompass the many complexities of
the nonlocal ion irradiation process within a familiar, intuitive
form, and seems to achieve good agreement with experiment. It therefore
demands reasons for the proposal of an alternative model.

\subsection{Differences with EBF: Fundamental Model}

We here present three concerns with the EBF model that are avoided
by the present approach.

\paragraph*{Surface-Dependent Body Force. }

The first, primary concern with the EBF model given by Eqn.(\ref{eqn: effective-body-force})
is its highly unusual suggestion that a ``body force'' could exist
that acts throughout the film, but which contains an explicit dependence
on the configuration of some nearby patch of surface. Traditionally,
true body forces that can act directly on bulk elements - also known
more descriptively as ``long-range forces'' - are completely independent
of nearby surfaces (since the latter are presumably much nearer than
the source of the long-range force). Any effects due to surfaces are
instead rigorously segregated into boundary conditions, to ensure
that the transmission of surface physics into the bulk is mediated
by the constitutive properties of the material itself. A formulation
like (\ref{eqn: effective-body-force}) blurs this traditional distinction,
suggesting the existence of a mechanism which depends on the surface,
but can nevertheless directly inform buried bulk parcels without any
consideration of the material in between.

\paragraph*{Functional Form of $\Psi$.}

A second concern is that, to obtain a good fit with experimental data,
the EBF model assumes that the function $\Psi$ take the form $\Psi=\cos\left(\theta-\gamma\right)$;
i.e., that the body force is proportional to the flux density at the
surface. This would be perfectly reasonable if the force $\mathbf{b}$
acted only at the surface - i.e., if it were a surface traction -
because it reflects the geometric dilution of the beam as the orientation
of the surface varies relative to the beam orientation. However, the
force $\mathbf{b}$ is instead proposed to act as a long-range force
throughout the bulk of the film, and as just discussed, true long-range
forces are typically approximately constant in any given location.
Hence, the correct form for the angle dependence of $\Psi$ is much
less obvious than it first appears, and the plausibility of the assumption
$\Psi=\cos\left(\theta-\gamma\right)$, upon which good agreement
with experiment rests, becomes less certain.

\paragraph*{Location of Stress.}

A final, minor concern is the inclusion of the undefined tensor $\mathbf{T}^{s}$
into the stress $\mathbf{T}$. True body forces are traditionally
not included within the stress tensor itself, but rather constitute
a separate item in the equations of conservation of momentum (\ref{eqn: conservation-of-momentum}).
The distinction is subtle: the resulting equations for the \emph{velocity
field} are the same in either case, but once velocities have been
calculated, the resulting \emph{stress field} (\ref{eqn: newton-plus-stress-CR})
depends qualitatively on the location of the effects of the beam.
So within the EBF model, if the effect of the beam is genuinely supposed
to appear within Eqn.(\ref{eqn: castro-constitutive}), then $\mathbf{T}^{s}$
must be rigorously defined because, as a part of the full stress $\mathbf{T}$,
it must play a role in the balance of forces at the free interface
$z=h\left(x,y\right)$. However, if the effect of the beam is truly
to be treated as a body force (as seems to be the intent), then it
should not appear in Eqn.(\ref{eqn: castro-constitutive}).

\paragraph*{Comparison with the present model.}

In contrast to these dilemmas inherent in the EBF model as currently
proposed, the model presented here maintains the traditional location
and form of terms, and makes fewer assumptions to achieve good agreement
with experiment. In particular, the ion beam acts within the bulk
constitutive law (\ref{eqn: otani-cr}) as a true tensor describing
stress-free anisotropic plastic flow. Acting as it does throughout
the entire body of the film, it is completely independent of the surface
configuration, as a traditional long-range mechanism should be. And,
as a true tensor, it is fully defined via Eqn.(\ref{eqn: APF-tensor})
and included in the force balance (\ref{eqn: surface-stress-BC})
at the free boundary. Finally, by starting at a level closer to the
microscopic process, our model based on the mechanism of anisotropic
plastic flow requires fewer, and weaker, assumptions. In particular,
the correct angular dependence of the wavelength on the irradiation
angle arises very naturally merely from the simple rotation of the
strain tensor with the beam.

\subsection{Differences with EBF: Theoretical Predictions}

In addition to fundamental differences at the level of modeling, the
present theory also differs from the EBF theory in terms of experimental
predictions.

\paragraph*{Steady State.}

The two models for stress produce different steady states. We first
observe that, in contrast to the constant pressure and linear flow
profile found here, Ref.\cite{castro-cuerno-ASS-2012} reports a depth-dependent
pressure and quadratic flow profile. This results in qualitatively
different scalings for the net flow through the film: $d^{2}$ vs.
$d^{3}$, which could be significant given the small size of $d$.
Next, combining Eqs.(4),(15-17) from Ref.\cite{castro-cuerno-ASS-2012},
one can show that the EBF model results in a steady stress throughout
the film of
\begin{equation}
\mathbf{T}_{EBF}=f_{E}\Psi\left(\theta\right)z\left[\begin{array}{cc}
\cos\left(\theta\right) & -\sin\left(\theta\right)\\
-\sin\left(\theta\right) & \cos\left(\theta\right)
\end{array}\right]+\mathbf{T}^{S},\label{eqn: castro-steady-stress}
\end{equation}
where the second component\textbf{ $\mathbf{T}^{S}$} remains undefined
within the EBF theory, and should potentially be omitted, as discussed
above. Focusing on first part associated with viscous stress, we observe
two differences with our Eqn.(\ref{eqn: steady-state}). First, the
magnitude of the viscous stress increases as the distance from the
free surface increases, reaching a maximum at the crystalline/amorphous
interface $z=-d$. However, considering that this stress is due to
incoming ions which are slowed by the film, one would expect the effect
of the beam to at most remain constant with film depth (as assumed
here in Eqn.(\ref{eqn: steady-state})), or, perhaps more realistically,
to decay with film depth. Second, Eqn.(\ref{eqn: castro-steady-stress})
has a different shape than our Eqn.(\ref{eqn: steady-state}), with
different angular dependence and principal stress directions, which
may be detectable using wafer curvature measurements. In particular,
when dotted with the upward-pointing unit vector $\hat{\mathbf{k}}$,
Eqn.(\ref{eqn: castro-steady-stress}) indicates the presence of stress
along the vertical direction; however, unless a body force is truly
present, the flat free boundary should allow all stress in the vertical
direction to relax to zero (as seen here in Eqn.(\ref{eqn: steady-state})).

\paragraph*{Wavelengths.}

A noteworthy success of the EBF theory is the apparently remarkable
agreement of its predictions on wavelength with experimental data;
our Figure~(\ref{fig: predicted-wavelength}) is nearly identical
to Figure 3 of Ref.\cite{castro-cuerno-ASS-2012}. However, it turns
out that the remarkable agreement with experiment depicted in that
figure were based on incorrect data. Taking the geometric mean of
MD estimates of 1.62 GPa \cite{kalyanasundaram-etal-MRC-2008}, and
preliminary experimental data of 200 MPa \cite{madi-etal-JPCM-2009},
the authors estimate the magnitude of the stress to be 569 MPa. This
value then leads to an estimate of $\left|\mathbf{f}_{E}\right|=0.424\,\mathrm{kg}/(\mathrm{nm}\,\mathrm{s})^{2}$,
producing the good fit with data. However, the lower, experimental
value of the stress was later found to be in error, and a more correct
value for Silicon irradiated by Argon at 250 eV has more recently
been revised to 1.4 GPa \cite{madi-thesis-2011}. Although the precise
means of calculating $\left|\mathbf{f}_{E}\right|$ are not specified
in Ref.\cite{castro-cuerno-ASS-2012}, if we assume that $\left|\mathbf{f}_{E}\right|\propto\left|\mathbf{T}_{0}\right|$
we may estimate a value associated with the revised stress of $\left|\mathbf{f}_{E}\right|\approx0.424\times1400\div569=1.04\,\mathrm{kg}/(\mathrm{nm}\,\mathrm{s})^{2}$,
which does not fit the data on wavelengths as well. Put another way,
the EBF model seems to over-estimate the effect of stress on growth
rate by a factor of about 2-3 relative to the current theory, which
provides better agreement with experiment when correct values of the
steady stress are used.

\paragraph*{Ripple Velocities.}

The theory of solid flow developed around the EBF model in Ref.\cite{castro-cuerno-ASS-2012}
predicts a ripple velocity of the form
\begin{equation}
V_{\text{ripple}}=\frac{d^{2}f_{E}\sin(2\theta)}{2\eta}\left(1-\frac{3}{8}Q^{2}\right)\label{eqn: castro-ripple-speed}
\end{equation}
(although the derivation appears to contain a few minor errors, such
as the use of the group velocity $\frac{d\omega}{dk}$ instead of
the phase velocity $\frac{\omega}{k}$ ). At first glance, Eqn.(\ref{eqn: castro-ripple-speed})
appears to have the same $\sin\left(2\theta\right)$ angle dependence
as our result \ref{eqn: ripple-velocity}; however, it also bears
a few important differences. First, within the parenthesis, Eqn.(\ref{eqn: castro-ripple-speed})
contains a constant term, and because it is typically assumed that
$Q\ll1$ (though see above), this term should be expected to dominate
the ripple velocity for large wavelengths. In contrast, our Eqn.(\ref{eqn: ripple-velocity})
contains no constant term which, because the critical value of $Q$
itself depends on $\theta$, should produce a different functional
form of the ripple velocity in terms of the angle. This difference
should be especially noticeable near the bifurcation at $45^{\circ}$,
where wavelengths become large: our model predicts stationary ripples
as $Q\to0$, whereas the EBF model predicts moving ripples. Second,
Eqn.(\ref{eqn: castro-ripple-speed}) depends on the ion-enhanced
viscosity of the material, but only weakly or not at all on the flux
(see Ref.\cite{castro-etal-arXiv-2012-solid-flow-drives} for a discussion
of parameter dependencies of $f_{E}$). In contrast, our Eqn.(\ref{eqn: ripple-velocity})
does not depend on the viscosity, and depends linearly on the flux,
consistent with the hypothesis that each impact induces a fixed amount
of anisotropic deformation in the material. These differences should
be testable experimentally with measurements of ripple velocity as
functions of various experimental parameters.

\section{Open Questions}

\paragraph*{Ripple Rotation.}

One of the fundamental successes of the original Bradley-Harper theory
of erosion was its ability to predict the oft-observed 90-degree rotation
of ripples at high angles of incidence \cite{bradley-harper-JVST-1988}.
Despite recent studies suggesting that this erosive effect is in fact
overwhelmed at most angles by the effect of redistributed atoms \cite{madi-etal-PRL-2011,norris-etal-NCOMM-2011},
no alternative to Bradley-Harper has been able to definitively predict
ripple rotation at high angles, and our theory is no exception. It
is likely that either erosive effects are still somehow dominant at
grazing angle, or that some other effect particular to grazing incidence
induces the ripple rotation. (We point out that Johnson et al. recently
predicted such a rotation based on molecular dynamics measurements
\cite{hossain-etal-APL-2011}. However, that prediction rests upon
apparent measurements of a \emph{negative} lateral momentum transfer
at grazing angles; we have instead observed lateral momentum transfer
to be positive at all angles \cite{norris-etal-NCOMM-2011}. Comparative
study of the methods used to obtain and filter these data would be
highly desirable.)

\paragraph*{Relationship to Mass Redistribution.}

A notable observation absent from the discussion surrounding the EBF
model is the observation that stress is actually the \emph{second}
distinct physical mechanism shown to produce reasonably good agreement
with experimentally-observed ripple wavelengths. It has been shown
previously that the leading order treatment of 'redistributed' target
atoms - those displaced from their original location but not sputtered
- produces second derivatives in the dispersion relation (\ref{eqn: longwave-dispersion-relation})
of exactly the same form as those reported here \cite{carter-vishnyakov-PRB-1996,davidovitch-etal-PRB-2007},
and molecular dynamics simulations suggest that these terms are of
just the right size to produce reasonably accurate predictions of
the observed ripple wavelengths \cite{norris-etal-NCOMM-2011}. Because
each effect, alone, seems able to predict the correct wavelengths,
the combination of both effects would be expected to be wrong.

One possible resolution to this dilemma appeals to the fact that stress,
ultimately, is generated by target atom displacements. Hence, morphology
evolution due to both target atom redistribution and stress are ultimately
driven solely by single ion impacts, just on different timescales.
So although they are properly distinct within a timescale-separated
framework, one could imagine the possibility of a generalized multi-scale
framework which incorporates both the displacements and stresses induced
by single ion impacts, as measured by molecular dynamics. In the process
of analyzing the contribution to stress (see, e.g., Refs.\cite{kalyanasundaram-etal-JEMT-2005,kalyanasundaram-etal-AM-2006,kalyanasundaram-etal-MRC-2008}),
the tensor $\mathbf{D}$ is something that could potentially be measured
as a function of angle, whereas the effective body force, until it
is connected directly with a microscopic mechanism, will likely be
more difficult to obtain.

A second possible explanation appeals to uncertainty surrounding the
exact value of the ion-enhanced viscosity $\eta$, which has only
been estimated using molecular dynamics. Whereas the wavelength predictions
in Ref.\cite{norris-etal-NCOMM-2011} due to mass redistribution relied
directly on $\eta$, within the present approach it was possible to
avoid estimating $\eta$ by instead measuring the steady stress $\left|\mathbf{T}_{0}\right|$,
which can be directly observed experimentally. Hence, while the accuracy
of the redistribution model is vulnerable to changes in estimates
of $\eta$, the accuracy of the anisotropic plastic flow model is
not. For this reason, the latter currently appears to this author
to be the more robust. So despite its technical problems, the authors
of the EBF model may ultimately be proved correct in their hypothesis
that ``Solid flow drives surface nano-patterning by ion-beam irradiation''
\cite{castro-etal-arXiv-2012-solid-flow-drives}.

\section{Conclusions}

In the pursuit of better understanding the role of stress in ion-irradiated
films, we have analyzed a model of stress based on the mechanism of
anisotropic plastic flow. This approach represents an alternative
to a recent ``effective body force'' (EBF) model of stress based
loosely on traditional fluid mechanics results. Based on distinct
microscopic mechanism, the model avoids several technical problems
in the EBF approach, and achieves remarkable experimental agreement
on ripple wavelengths using more accurate experimental parameters
and fewer underlying assumptions. In addition, it makes more intuitive
predictions on the steady stress, and alternate, testable predictions
on the ripple velocity. Although challenges remain, notably a failure
to predict ripple rotation at grazing incidence angles, and a need
to distinguish stress effects from those of redistributed atoms, this
closer link to physical origins, and expanded predictive power for
Argon-irradiated Silicon, make the model a potentially valuable stepping
stone toward the goal of a predictive theory with the quantitative
accuracy necessary for industrial relevance.

 \clearpage{}

\bibliographystyle{unsrt}
\bibliography{/home/snorris/Dropbox/research/bibliography/tagged-bibliography}

\begin{thebibliography}{10}

\bibitem{chan-chason-JAP-2007}
W.~L. Chan and E.~Chason.
\newblock Making waves: kinetic processes controlling surface evolution during
  low energy ion sputtering.
\newblock {\em J. Appl. Phys.}, 101:121301, 2007.

\bibitem{frost-etal-APA-2008}
F.~Frost, B.~Ziberi, A.~Schindler, and B.~Rauschenbach.
\newblock Surface engineering with ion beams: From self-organized
  nanostructures to ultra-smooth surfaces.
\newblock {\em Appl. Phys. A}, 91:551--559, 2008.

\bibitem{facsko-etal-SCIENCE-1999}
S.~Facsko, T.~Dekorsy, C.~Koerdt, C.~Trappe, H.~Kurz, A.~Vogt, and H.~L.
  Hartnagel.
\newblock Formation of ordered nanoscale semiconductor dots by ion sputtering.
\newblock {\em Science}, 285:1551--1553, 1999.

\bibitem{ozaydin-etal-APL-2005}
G.~Ozaydin, A.S. Ozcan, Y.Y. Wang, K.F. Ludwig, H.~Zhou, R.L. Headrick, and
  D.P. Siddons.
\newblock Real-time x-ray studies of {Mo}-seeded {Si} nanodot formation during
  ion bombardment.
\newblock {\em Applied Physics Letters}, 87:163104, 2005.

\bibitem{macko-etal-NanoTech-2010}
S.~Macko, F.~Frost, B.~Ziberi, D.F. Forster, and T.~Michely.
\newblock Is {keV} ion-induced pattern formation on {Si(001)} caused by metal
  impurities?
\newblock {\em Nanotechnology}, 21:085301, 2010.

\bibitem{bradley-shipman-PRL-2010}
R.~Mark Bradley and Patrick~D. Shipman.
\newblock Spontaneous pattern formation induced by ion bombardment of binary
  compounds.
\newblock {\em Physical Review Letters}, 105:145501, 2010.

\bibitem{gago-etal-APL-2006}
R.~Gago, L.~V{\'a}squez, O.~Plantevin, T.~H. Metzger, J.~Mu{\~n}oz-Garc{\'i}a,
  R.~Cuerno, and M.~Castro.
\newblock Order enhancement and coarsening of self-organized silicon nanodot
  patterns induced by ion-beam sputtering.
\newblock {\em Appl. Phys. Lett.}, 89:233101, 2006.

\bibitem{ziberi-etal-APL-2008}
B.~Ziberi, F.~Frost, M.~Tartz, H.~Neumann, and B.~Rauschenbach.
\newblock Ripple rotation, pattern transitions, and long range ordered dots on
  silicon by ion beam erosion.
\newblock {\em Appl. Phys. Lett.}, 92:063102, 2008.

\bibitem{madi-etal-2008-PRL}
C.~S. Madi, B.~P. Davidovitch, H.~B. George, S.~A. Norris, M.~P. Brenner, and
  M.~J. Aziz.
\newblock Multiple bifurcation types and the linear dynamics of ion sputtered
  surfaces.
\newblock {\em Phys. Rev. Lett.}, 101:246102, 2008.

\bibitem{facsko-etal-PRB-2004}
S.~Facsko, T.~Bobek, A.~Stahl, and H.~Kurz.
\newblock Dissipative continuum model for self-organized pattern formation
  during ion-beam erosion.
\newblock {\em Phys. Rev. B}, 69:153412, 2004.

\bibitem{davidovitch-etal-PRB-2007}
B.~P. Davidovitch, M.~J. Aziz, and M.~P. Brenner.
\newblock On the stabilization of ion sputtered surfaces.
\newblock {\em Phys. Rev. B}, 76:205420, 2007.

\bibitem{bradley-PRB-2011a}
R.M. Bradley.
\newblock Redeposition of sputtered material is a nonlinear effect.
\newblock {\em Physical Review B}, 83:075404, 2011.

\bibitem{norris-PRB-2012-viscoelastic-normal}
S.~A. Norris.
\newblock Stability analysis of a viscoelastic model for ion-irradiated
  silicon.
\newblock {\em Physical Review B}, 85:155325, 2012.

\bibitem{madi-aziz-ASS-2012}
Charbel~S. Madi and Michael~J. Aziz.
\newblock Multiple scattering causes the low energy-low angle constant
  wavelength bifurcation of argon ion bombarded silicon surfaces.
\newblock {\em Applied Surface Science}, 258:4112--4115, 2012.
\newblock (IINM2011 Bhubaneswar Conference Proceedings).

\bibitem{ozaydin-etal-JVSTB-2008}
G.~Ozaydin, Jr. K.~F.~Ludwig, H.~Zhou, and R.~L. Headrick.
\newblock Effects of mo seeding on the formation of si nanodots during
  low-energy ion bombardment.
\newblock {\em J. Vac. Sci. Technol. B}, 26:551, 2008.

\bibitem{ozaydin-ludwig-JPCM-2009}
G.~Ozaydin-Ince and K.~F. Ludwig~Jr.
\newblock In situ x-ray studies of native and mo-seeded surface nanostructuring
  during ion bombardment of si(100).
\newblock {\em J. Phys. Cond. Matt.}, 21:224008, 2009.

\bibitem{zhang-etal-NJoP-2011}
Kun Zhang, Marc Br{\"o}tzmann, and Hans Hofs{\"a}ss.
\newblock Surfactant-driven self-organized surface patterns by ion beam
  erosion.
\newblock {\em New Journal of Physics}, 13:013033, 2011.

\bibitem{shenoy-chan-chason-PRL-2007}
V.~B. Shenoy, W.~L. Chan, and E.~Chason.
\newblock Compositionally modulated ripples induced by sputtering of alloy
  surfaces.
\newblock {\em Physical Review Letters}, 98:256101, 2007.

\bibitem{shipman-bradley-PRB-2011}
P.~D. Shipman and R.~M. Bradley.
\newblock Theory of nanoscale pattern formation induced by normal-incidence ion
  bombardment of binary compounds.
\newblock {\em Physical Review B}, 84:085420, 2011.

\bibitem{bradley-PRB-2011c}
R.~Mark Bradley.
\newblock Theory of nanodot and sputter cone arrays produced by ion sputtering
  with concurrent deposition of impurities.
\newblock {\em Physical Review B}, 83:195410, 2011.

\bibitem{bradley-JAP-2012}
R.~M. Bradley.
\newblock Surface instability of binary compounds caused by sputter yield
  amplification.
\newblock {\em Journal of Applied Physics}, in press, 2012.

\bibitem{abrasonis-morawetz-PRB-2012}
G.~Abrasonis and K.~Morawetz.
\newblock Instability types at ion-assisted alloy deposition: from
  two-dimensional to three-dimensional nanopattern growth.
\newblock {\em Physical Review B}, submitted, 2012.
\newblock arXiv:1109.5461v2.

\bibitem{norris-PRB-2012-chemical-instability}
Scott~A. Norris.
\newblock A chemically-driven finite-wavelength instability in ion-irradiated
  compound semiconductors.
\newblock arXiv:1205.6834v1 [cond-mat.mtrl-sci].

\bibitem{sigmund-PR-1969}
P.~Sigmund.
\newblock Theory of sputtering. {I}. {S}puttering yield of amorphous and
  polycrystalline targets.
\newblock {\em Phys. Rev.}, 184:383--416, 1969.

\bibitem{sigmund-JMS-1973}
P.~Sigmund.
\newblock A mechanism of surface micro-roughening by ion bombardment.
\newblock {\em J. Mater. Sci.}, 8:1545--1553, 1973.

\bibitem{bradley-harper-JVST-1988}
R.~M. Bradley and J.~M.E. Harper.
\newblock Theory of ripple topography induced by ion bombardment.
\newblock {\em J. Vac. Sci. Technol.}, 6:2390--2395, 1988.

\bibitem{carter-vishnyakov-PRB-1996}
G.~Carter and V.~Vishnyakov.
\newblock Roughening and ripple instabilities on ion-bombarded si.
\newblock {\em Phys. Rev. B}, 54:17647--17653, 1996.

\bibitem{volkert-JAP-1991}
C.~A. Volkert.
\newblock Stress and plastic flow in silicon during amorphization by ion
  bombardment.
\newblock {\em J. Appl. Phys.}, 70:3521, 1991.

\bibitem{umbach-etal-PRL-2001}
C.~C. Umbach, R.~L. Headrick, and K.-C. Chang.
\newblock Spontaneous nanoscale corrugation of ion-eroded $\mathrm{SiO}_2$: The
  role of ion-irradiation-enhanced viscous flow.
\newblock {\em Phys. Rev. Lett.}, 87:246104, 2001.

\bibitem{cuerno-etal-NIMB-2011}
R.~Cuerno, M.~Castro, J.~Mu{\~n}oz-Garc{\'i}a, R.~Gago, and L.~V{\'a}squez.
\newblock Nanoscale pattern formation at surfaces under ion-beam sputtering:
  {A} perspective from continuum models.
\newblock {\em Nuclear Instruments and Methods in Physics Research B},
  269:894--900, 2011.

\bibitem{castro-cuerno-ASS-2012}
Mario Castro and Rodolfo Cuerno.
\newblock Hydrodynamic approach to surface pattern formation by ion beams.
\newblock {\em Applied Surface Science}, 258:4171--4178, 2012.

\bibitem{castro-etal-arXiv-2012-solid-flow-drives}
M.~Castro, R.~Gago, L.~V{\'a}squez, J.~Mu{\~n}oz-Garc{\'i}a, and R.~Cuerno.
\newblock Solid flow drives surface nanopatterning by ion-beam irradiation.
\newblock arXiv:1203.1167v1 [cond-mat.mtrl-sci].

\bibitem{oron-davis-bankoff-RMP-1997}
Alexander Oron, Stephen~H. Davis, and S.~George Bankoff.
\newblock Long-scale evolution of thin liquid films.
\newblock {\em Reviews of Modern Physics}, 69:931--980, 1997.

\bibitem{trinkaus-ryazanov-PRL-1995-viscoelastic}
H.~Trinkaus and A.~I. Ryazanov.
\newblock Viscoelastic model for the plastic flow of amorphous solids under
  energetic ion bombardment.
\newblock {\em Physical Review Letters}, 75:5072--5075, 1995.

\bibitem{trinkaus-NIMB-1998-viscoelastic}
H.~Trinkaus.
\newblock Dynamics of viscoelastic flow in ion tracks: origin of plastic
  deformation of amorphous materials.
\newblock {\em Nuclear Instruments and Methods in Physics Research B},
  146:204--216, 1998.

\bibitem{snoeks-etal-JAP-1995}
E.~Snoeks, T.~Weber, A.~Cacciato, and A.~Polman.
\newblock {MeV} ion irradiation-induced creation and relaxation of mechanical
  stress in silica.
\newblock {\em J. Appl. Phys.}, 78:4723, 1995.

\bibitem{van-dillen-etal-APL-2001-colloidal-ellipsoids}
T.~van Dillen, A.~Polman, W.~Fukarek, and A.~van Blaaderen.
\newblock Energy-dependent anisotropic deformation of colloidal silica
  particles under mev au irradiation.
\newblock {\em Applied Physics Letters}, 78:910--912, 2001.

\bibitem{van-dillen-etal-PRB-2005-viscoelastic-model}
T.~van Dillen, A.~Polman, P.~R. Onck, and E.~van~der Giessen.
\newblock Anisotropic plastic deformation by viscous flow in ion tracks.
\newblock {\em Physical Review B}, 71:024103, 2005.

\bibitem{otani-etal-JAP-2006}
K.~Otani, X.~Chen, J.~W. Hutchinson, J.~F. Chervinsky, and M.~J. Aziz.
\newblock Three-dimensional morphology evolution of $\mathrm{SiO}_2$ patterned
  films under {MeV} ion irradiation.
\newblock {\em J. Appl. Phys.}, 100:023535, 2006.

\bibitem{van-dillen-etal-APL-2003-colloidal-ellipsoids}
T.~van Dillen, A.~Polman, C.~M. van Kats, and A.~van Blaaderen.
\newblock Ion beam-induced anisotropic plastic deformation at 300 kev.
\newblock {\em Applied Physics Letters}, 83:4315--4317, 2003.

\bibitem{kim-etal-JAP-2006-stressed-keV-films}
Y.-R. Kim, P.~Chen, M.~J. Aziz, D.~Branton, and J.~J. Vlassak.
\newblock Focused ion beam induced deflections of freestanding thin films.
\newblock {\em Journal of Applied Physics}, 100:104322, 2006.

\bibitem{george-etal-JAP-2010}
H.~B. George, Y.~Tang, X.~Chen, J.~Li, J.~W. Hutchinson, J.~A. Golovchenko, and
  M.~J. Aziz.
\newblock Nanopore fabrication in amorphous si: Viscous flow model and
  comparison to experiment.
\newblock {\em Journal of Applied Physics}, 108:014310, 2010.

\bibitem{norris-etal-NCOMM-2011}
S.~A. Norris, J.~Samela, L.~Bukonte, M.~Backman, D.~F.~K. Nordlund, C.S. Madi,
  M.P. Brenner, and M.J. Aziz.
\newblock Molecular dynamics of single-particle impacts predicts phase diagrams
  for large scale pattern formation.
\newblock {\em Nature Communications}, 2:276, 2011.

\bibitem{malvern-1977}
Lawrence~E. Malvern.
\newblock {\em Introduction to the Mechanics of a Continuous Medium}.
\newblock Prentice Hall, 1977.
\newblock ISBN: 0134876032.

\bibitem{moseler-etal-SCIENCE-2005}
M.~Moseler, P.~Gumbsch, C.~Casiraghi, A.~C. Ferrari, and J.~Robertson.
\newblock The ultrasmoothness of diamond-like carbon surfaces.
\newblock {\em Science}, 309:1545--1548, 2005.

\bibitem{kalyanasundaram-etal-APL-2008}
N.~Kalyanasundaram, M.~Ghazisaeidi, J.~B. Freund, and H.~T. Johnson.
\newblock Single impact crater functions for ion bombardment of silicon.
\newblock {\em Appl. Phys. Lett.}, 92:131909, 2008.

\bibitem{kalyanasundaram-etal-AM-2006}
N.~Kalyanasundaram, M.~C. Moore, J.~B. Freund, and H.~T. Johnson.
\newblock Stress evolution due to medium-energy ion bombardment of silicon.
\newblock {\em Acta Materialia}, 54:483--491, 2006.

\bibitem{madi-thesis-2011}
C.~S. Madi.
\newblock {\em Linear Stability and Instability Patterns in Ion Bombarded
  Silicon Surfaces}.
\newblock PhD thesis, Harvard University, 2011.

\bibitem{madi-2009-MRS-castro-response}
C.~S. Madi and M.~Aziz.
\newblock private communication.

\bibitem{orchard-ASR-1962}
S.~E. Orchard.
\newblock On surface levelling in viscous liquids and gels.
\newblock {\em Appl. Sci. Res.}, 11A:451, 1962.

\bibitem{madi-etal-JPCM-2009}
C.~S. Madi, H.~B. George, and M.~J. Aziz.
\newblock Linear stability and instability patterns in ion-sputtered silicon.
\newblock {\em J. Phys. Cond. Matt.}, 21:224010, 2009.

\bibitem{vauth-mayr-PRB-2007}
S.~Vauth and S.~G. Mayr.
\newblock Relevance of surface viscous flow, surface diffusion, and ballistic
  effects in kev ion smoothing of amorphous surfaces.
\newblock {\em Phys. Rev. B}, 75:224107, 2007.

\bibitem{eaglesham-etal-PRL-1993}
D.~J. Eaglesham, A.~E. White, L.~C. Feldman, N.~Moriya, and D.~C. Jacobson.
\newblock Equilibrium shape of {Si}.
\newblock {\em Phys. Rev. Lett.}, 70:1643, 1993.

\bibitem{alkemade-PRL-2006}
P.~F.~A. Alkemade.
\newblock Propulsion of ripples on glass by ion bombardment.
\newblock {\em Phys. Rev. Lett.}, 96:107602, 2006.

\bibitem{gnaser-etal-NIMB-2012-ripple-speed}
Hubert Gnaser, Bernhard Reuscher, and Anna Zeuner.
\newblock Propagation of nanoscale ripples on ion-irradiated surfaces.
\newblock {\em Nuclear Instruments and Methods in Physics Research B},
  285:142--147, 2012.

\bibitem{kalyanasundaram-etal-MRC-2008}
Nagarajan Kalyanasundaram, Molly Wood, Jonathan~B. Freund, and H.T. Johnson.
\newblock Stress evolution to steady state in ion bombardment of silicon.
\newblock {\em Mechanics Research Communications}, 35:50--56, 2008.

\bibitem{madi-etal-PRL-2011}
C.~S. Madi, E.~Anzenberg, K.~F. Ludwig~Jr., , and M.~J. Aziz.
\newblock Mass redistribution causes the structural richness of ion-irradiated
  surfaces.
\newblock {\em Phys. Rev. Lett.}, 106:066101, 2011.

\bibitem{hossain-etal-APL-2011}
M.~Z. Hossain, K.~Das, J.~B. Freund, and H.~T. Johnson.
\newblock Ion impact crater asymmetry determines surface ripple orientation.
\newblock {\em Applied Physics Letters}, 99:151913, 2011.

\bibitem{kalyanasundaram-etal-JEMT-2005}
N.~Kalyanasundaram, J.~B. Freund, and H.~T. Johnson.
\newblock Atomistic determination of continuum mechanical properties of
  ion-bombarded silicon.
\newblock {\em Journal of Engineering Materials and Technology}, 127:457--461,
  2005.

\end{thebibliography}

\clearpage{}

\appendix

\section{Solution of the Linearized Equations}

In this Appendix we present the details of the calculation by which
the dispersion relation (\ref{eqn: full-dispersion-relation}) was
obtained.

\paragraph*{General solution from the bulk Equations.}

In the linear ansatz (\ref{eqn: perturbations}), the (already-linear)
Stokes equations become
\begin{eqnarray}
-ik_{1}p\left(z\right)+\eta\left[-\left(k_{1}^{2}+k_{2}^{2}\right)u+u^{\prime\prime}\right] & = & 0\label{eqn: app-u}\\
-ik_{2}p\left(z\right)+\eta\left[-\left(k_{1}^{2}+k_{2}^{2}\right)v+v^{\prime\prime}\right] & = & 0\label{eqn: app-v}\\
-p^{\prime}\left(z\right)+\eta\left[-\left(k_{1}^{2}+k_{2}^{2}\right)w+w^{\prime\prime}\right] & = & 0\label{eqn: app-w}\\
ik_{1}u+ik_{2}v+w^{\prime} & = & 0\label{eqn: app-uvw}
\end{eqnarray}
The first step in the solution is to eliminate $u,v,w$ in the following
manner:
\begin{enumerate}
\item differentiate Eqn.(\ref{eqn: app-uvw}), to get an expression for
$w^{\prime\prime}$
\item insert the resulting expression for $w^{\prime\prime}$ into Eqn.(\ref{eqn: app-w})
\item differentiate the result to get an expression for $p^{\prime\prime}\left(z\right)$
\item in the resulting expression, eliminate $w^{\prime}$ using (\ref{eqn: app-uvw})
\item finally, eliminate $u$ and $v$ using (\ref{eqn: app-u}) and (\ref{eqn: app-v})
\end{enumerate}
The resulting homogeneous equation for $p\left(z\right)$ is 
\[
p^{\prime\prime}-\left(k_{1}^{2}+k_{2}^{2}\right)p=0
\]
with solutions
\begin{equation}
p\left(z\right)=\bar{A}\cosh\left(Rz\right)+\bar{B}\sinh\left(Rz\right),\label{eqn: app-p-sol}
\end{equation}
where $R=\sqrt{k_{1}^{2}+k_{2}^{2}}$.

From here, we simply insert the solution (\ref{eqn: app-p-sol}) back
into (\ref{eqn: app-u})-(\ref{eqn: app-w}) to get inhomogeneous
equations for $u,v,w$: these have solutions
\begin{eqnarray}
u & = & C\cosh\left(Rz\right)+D\sinh\left(Rz\right)+\frac{ik_{1}}{2\eta R}z\left[\bar{B}\cosh\left(Rz\right)+\bar{A}\sinh\left(Rz\right)\right]\label{eqn: app-u-sol}\\
v & = & E\cosh\left(Rz\right)+F\sinh\left(Rz\right)+\frac{ik_{2}}{2\eta R}z\left[\bar{B}\cosh\left(Rz\right)+\bar{A}\sinh\left(Rz\right)\right]\label{eqn: app-v-sol}\\
w & = & G\cosh\left(Rz\right)+H\sinh\left(Rz\right)+\frac{1}{2\eta}z\left[\bar{A}\cosh\left(Rz\right)+\bar{B}\sinh\left(Rz\right)\right].\label{eqn: app-w-sol}
\end{eqnarray}
At this point, however, we have too many constants because we first
differentiated Eqn.(\ref{eqn: app-uvw}), raising its order. Re-considering
Eqn.(\ref{eqn: app-uvw}) in its original form, we insert (\ref{eqn: app-u-sol})-(\ref{eqn: app-w-sol})
and collect like terms in $\sinh\left(Rz\right)$ and $\cosh\left(Rz\right)$
to obtain
\begin{eqnarray*}
\frac{ik_{1}}{R}D+\frac{ik_{2}}{R}F+\frac{1}{2\eta R}\bar{B}+G & = & 0\\
\frac{ik_{1}}{R}C+-\frac{ik_{2}}{R}E+\frac{1}{2\eta R}\bar{A}+H & = & 0.
\end{eqnarray*}
From here, we choose to replace $\bar{A}$ and $\bar{B}$ with the
other constants to obtain the following expressions for the pressure
and velocity: 
\begin{equation}
\begin{aligned}p_{1}\left(z\right) & =-2\eta\left[\left(RH+ik_{1}C+ik_{2}E\right)\cosh\left(Rz\right)+\left(RG+ik_{1}D+ik_{2}F\right)\sinh\left(Rz\right)\right]\\
u_{1}\left(z\right) & =C\cosh\left(Rz\right)+D\sinh\left(Rz\right)-\frac{ik_{1}}{R}\left[\begin{array}{c}
\left(RG+ik_{1}D+ik_{2}F\right)z\cosh\left(Rz\right)\\
+\left(RH+ik_{1}C+ik_{2}E\right)z\sinh\left(Rz\right)
\end{array}\right]\\
v_{1}\left(z\right) & =E\cosh\left(Rz\right)+F\sinh\left(Rz\right)-\frac{ik_{2}}{R}\left[\begin{array}{c}
\left(RG+ik_{1}D+ik_{2}F\right)z\cosh\left(Rz\right)\\
+\left(RH+ik_{1}C+ik_{2}E\right)z\sinh\left(Rz\right)
\end{array}\right]\\
w_{1}\left(z\right) & =G\cosh\left(Rz\right)+H\sinh\left(Rz\right)-\left[\begin{array}{c}
\left(RG+ik_{1}D+ik_{2}F\right)z\sinh\left(Rz\right)\\
+\left(RH+ik_{1}C+ik_{2}E\right)z\cosh\left(Rz\right)
\end{array}\right]
\end{aligned}
,\label{eqn: linear-solution}
\end{equation}
where $R=\sqrt{k_{1}^{2}+k_{2}^{2}}$, and $C,D,E,F,G,H$ are integration
constants.

\paragraph*{Integration constants from Boundary Conditions.}

To find the integration constants, we apply the linearized versions
of the three no penetration/slip conditions (\ref{eqn: no-slip-or-penetration-BC})
at the amorphous/crystalline interface $z=h_{1}\left(x,y\right)$,
and the three stress balance conditions (\ref{eqn: surface-stress-BC})
at the free interface $z=h_{0}+h_{1}\left(x,y\right)$. At the bottom
of the film, $z=g_{1}\left(x,y\right)=h_{1}\left(x,y\right)$, the
no-slip boundary condition (\ref{eqn: no-slip-or-penetration-BC})
linearizes to the related condition
\begin{equation}
\mathbf{v}_{1}\left(x,y,0\right)+\frac{\partial\mathbf{v}_{0}}{\partial z}\left(x,y,0\right)\cdot h_{1}=0\label{eqn: linearized-bottom}
\end{equation}
at $z=0$, giving
\begin{equation}
\begin{aligned}C & =-3fA\sin\left(2\theta\right)h_{1}\left(x,y\right)\\
E & =0\\
G & =0
\end{aligned}
.\label{eqn: CEG}
\end{equation}
At the top of the film, $z=h_{0}+h_{1}\left(x,y\right)$, we apply
the stress condition (\ref{eqn: surface-stress-BC}); again, this
linearizes to a related condition 
\[
\mathbf{T}_{0}\cdot\hat{\mathbf{n}}_{1}+\mathbf{T}_{1}\cdot\hat{\mathbf{n}}_{0}=-\gamma\kappa_{1}\hat{\mathbf{n}}_{0}
\]
at $z=h_{0}$, which forms a matrix equation for the remaining constants$\left\{ D,\, F,\, H\right\} $:
\begin{equation}
\left[\begin{array}{ccc}
R\bar{C}+\frac{k_{1}^{2}}{R}\left(\bar{C}+2Q\bar{S}\right) & \frac{k_{1}k_{2}}{R}\left(\bar{C}+2Q\bar{S}\right) & -2ik_{1}Q\bar{C}\\
\frac{k_{1}k_{2}}{R}\left(\bar{C}+2Q\bar{S}\right) & R\bar{C}+\frac{k_{2}^{2}}{R}\left(\bar{C}+2Q\bar{S}\right) & -2ik_{2}Q\bar{C}\\
-2ik_{1}Q\bar{C} & -2ik_{2}Q\bar{C} & 2R\left(\bar{C}-Q\bar{S}\right)
\end{array}\right]\left[\begin{array}{c}
D\\
F\\
H
\end{array}\right]=\left[\begin{array}{c}
\bar{\alpha}\\
\bar{\beta}\\
\bar{\gamma}
\end{array}\right],\label{eqn: matrix-for-coeffs}
\end{equation}
where $Q=h_{0}R$, $\bar{C}=\cosh\left(Q\right)$, $\bar{S}=\sinh\left(Q\right)$,
and 
\begin{equation}
\left[\begin{array}{c}
\bar{\alpha}\\
\bar{\beta}\\
\bar{\gamma}
\end{array}\right]=-\gamma\left[\begin{array}{c}
0\\
0\\
R^{2}
\end{array}\right]+6\eta fAh_{1}\left[\begin{array}{c}
ik_{1}\cos\left(2\theta\right)\\
ik_{2}\cos^{2}\left(\theta\right)\\
0
\end{array}\right]+3\eta fAh_{1}\sin\left(2\theta\right)\left[\begin{array}{c}
R\bar{S}+\frac{k_{1}^{2}}{R}\left(\bar{S}+2Q\bar{C}\right)\\
\frac{k_{1}k_{2}}{R}\left(\bar{S}+2Q\bar{C}\right)\\
-2ik_{1}Q\sinh\left(Q\right)
\end{array}\right].\label{eqn: abbbcb}
\end{equation}
The three column vectors on the right hand side of Eqn.(\ref{eqn: abbbcb})
represent, respectively, the effects of surface tension, the beam
stress acting in the bulk, and the effect under that stress of the
non-planar bottom boundary. 

The solution to Eqn.(\ref{eqn: matrix-for-coeffs}) can be found using
Cramer's Rule or a computer algebra system, yielding the result
\begin{equation}
\begin{aligned}D & =\frac{1}{\Delta}\left[2\bar{\alpha}R^{2}\left(\bar{C}^{2}-Q\bar{S}\bar{C}\right)+2k_{2}\left(\bar{\alpha}k_{2}-\bar{\beta}k_{1}\right)\left(\bar{C}^{2}+Q\bar{S}\bar{C}+2Q^{2}\right)+2i\bar{\gamma}k_{1}RQ\bar{C}^{2}\right]\\
F & =\frac{1}{\Delta}\left[2\bar{\beta}R^{2}\left(\bar{C}^{2}-Q\bar{S}\bar{C}\right)-2k_{1}\left(\bar{\alpha}k_{2}-\bar{\beta}k_{1}\right)\left(\bar{C}^{2}+Q\bar{S}\bar{C}+2Q^{2}\right)+2i\bar{\gamma}k_{2}RQ\bar{C}^{2}\right]\\
H & =\frac{1}{\Delta}\left[2iR\left(\bar{\alpha}k_{1}+\bar{\beta}k_{2}\right)Q\bar{C}^{2}+2\bar{\gamma}R^{2}\left(\bar{C}^{2}+Q\bar{S}\bar{C}\right)\right]
\end{aligned}
,\label{eqn: DFH}
\end{equation}
where $\bar{\alpha},$ $\bar{\beta}$, $\bar{\gamma}$ remain as given
in Eqn.(\ref{eqn: abbbcb}) and
\[
\Delta=2R^{3}\cosh\left(Q\right)\left[1+2Q^{2}+\cosh\left(2Q\right)\right]
\]
is the determinant of the matrix on the left hand side of Eqn.(\ref{eqn: matrix-for-coeffs}).

\paragraph*{Dispersion Relation from the Kinematic Condition.}

Having uniquely determined the pressure and velocity fields, it remains
to apply the kinematic condition (\ref{eqn: kinematic-condition}),
which linearizes to

\begin{equation}
\sigma\left(k_{1},k_{2}\right)=w_{1}\left(h_{0}\right)-u_{0}\left(h_{0}\right)\frac{\partial h_{1}}{\partial x}-v_{0}\left(h_{0}\right)\frac{\partial h_{1}}{\partial y}.\label{eqn: linearized-kinematic-condition}
\end{equation}
From the general solution (\ref{eqn: linear-solution}) in the main
text, and the solutions (\ref{eqn: CEG}) and (\ref{eqn: DFH}) for
the integration constants, we can evaluate $w_{1}\left(h_{0}\right)$,
which leads directly to the dispersion relation (\ref{eqn: full-dispersion-relation})
in the main text.
\end{document}